**Aging, Fragility and Reversibility Window in Bulk Alloy Glasses**


*S. Chakravarty, D.G. Georgiev, P.Boolchand*

Department of ECECS, University of Cincinnati, Cincinnati, OH 45221-0030,USA

*M.Micoulaut*
Laboratoire de Physique Theorique des Liquides, Universite Pierre et Marie Curie,
Boite 121, 4 Place Jussieu, 75252 Paris, Cedex 05, France



Non-reversing relaxation enthalpies ($\Delta H_{nr}$) at glass transitions $T_g(x)$ in the $P_x Ge_x Se_{1-2x}$ ternary display a wide, sharp and deep global minimum (~0) in the $0.09 < x < 0.145$ range, within which $T_g$ becomes *thermally reversing*. In the *reversibility window* these glasses are found *not to age*, in contrast to *aging* observed for fragile glass compositions outside the *window*. *Thermal reversibility* and *lack of aging* are paradigms that molecular glasses in the *window* share with *proteins* in *transition states,* which result from structural self-organization in both systems. In proteins the self-organized structures appear to be at places where life sustaining repeating foldings and unfoldings occur.


Aging occurs in many materials, both organic and inorganic. Inorganic crystals age under electrical, mechanical, or thermal stresses, often as a result of dislocation motion. Aging in organic materials is more complex, occurring as hydrogen bonding





configurations are altered as a result of thermal cycling. Here we will show that inorganic non-crystalline nanonetworks are universally divided into three regimes of composition, two of which age rapidly, while the third regime scarcely ages at all. The third regime defines a narrow window of composition that appears to have much in common mechanically with selected organic nanonetworks, namely the polypeptide chains that form proteins. Aging is not evident in data obtained by conventional structural methods (diffraction) but it is measured very accurately by modulated scanning calorimetry.

New ideas on the nature of glass transitions ( $T_g$ ) have emerged in recent years from examination[1-3] of the *non-reversing relaxation enthalpy* ($\Delta H_{nr}$) associated with $T_g$. The enthalpy is a signature of ergodicity breaking events when structural arrest of a glass forming liquid occurs near $T_g$. Examined as a function of mean coordination number *r* of network glasses, the endotherm ($\Delta H_{nr}$) is found to nearly vanish[1-5] across *compositional windows, $r_c(1) < r < r_c(2)$,* within which glass transition becomes *thermally reversing*. Furthermore, these *reversibility windows* are found to be closely related to variations in Raman optical elasticities[3-5]. There are distinct elastic power-laws[3-5], for glasses in the three regions: $r < r_c(1)$, *r* in between $r_c(1)$ and $r_c(2)$), and $r > r_c(2)$, which have been observed in chalcogenide glasses. Using the idea of Lagrangian constraints[6], as well as graph theory[7] and also numerical simulations[7,8], J.C.Phillips and M.F. Thorpe have identified the existence of *three* generic elastic phases as a function of *r*; *floppy, intermediate and stressed rigid*. Thus, $r_c(1)$ and $r_c(2)$, mark the *onset* and *end* of the *reversibility window*, also represent the two *phase boundaries* between these *three elastic*





*phases*. In the *reversibility windows*, ($r_c(1) < r < r_c(2)$) the glasses are in *intermediate phases*[3,8,9] between *floppy* ($r < r_c(1)$) and *stressed rigid* ($r > r_c(2)$) ones, and melt *reversibly* ($\Delta H_{nr} \sim 0$) at $T_g$. Intermediate phases are generally centered on mean coordination numbers close to 2.40, and are described as elastically isostatic (rigid but unstressed). Here we connect these mechanical properties to the rate of network aging, as measured by changes in the kinetics of the glass transition in samples relaxed at room temperature (far below the glass transition temperatures) over several months.

$T_g$'s are an intimate measure of network connectedness ($r$) as demonstrated[10] by stochastic agglomeration theory (SAT). The structural interpretation of $T_g$ has found quantitative support in the remarkable agreement[1-5,11] between measured and SAT predicted variations of $T_g(r)$ in alloyed bulk glasses of the oxides and chalcogenides. In several binary selenides, ($T$ or $Pn$)$_x$Se$_{1-x}$, where $T$ is a tathogen (Si, Ge) and $Pn$ a pnictide ( P,As), trends in $T_g(x)$ display global maxima near chemical thresholds[12] ( Fig. 1a). In ternary selenides, $T_x Pn_x Se_{1-x}$, containing *equal fractions* of $T$ and $Pn$ atoms, these global maxima are conspicuously absent[12] however, and $T_g(x)$ is found to increase monotonically with x , as illustrated for the case of $T$ = Ge, $Pn$ = P ternary in Fig.1a. These contrasting variations, are suggestive[11] of *nanoscale phase separation* (*nsps*) of backbones in the binaries but their absence in the ternaries, making the latter systems especially attractive to probe connectivity related phase transitions. In this Letter, we identify the r*eversibility window* (0.09 < x < 0.145) in the P$_x$Ge$_x$Se$_{1-2x}$ ternary and find it to be wide, sharp and deep. Furthermore, glasses *in the window* are found not to age, a behavior that is in sharp contrast to aging observed for glass compositions *outside* the





window. *Thermal reversibility and absence of aging* represent functions that are common to both *glasses* in *reversibility windows* and *proteins in folding states*[13,14], and lead to *self-organization* of the former and *sustenance of life* in the latter.

The starting materials, 99.999% Ge, P and Se from Cerac Inc., were handled in a dry nitrogen ambient and reacted in evacuated ( $5 \times 10^{-7}$ Torr) fused quartz ampoules of 5mm id. Ampoules were heated slowly to 950ºC in a rotating furnace and held at that T for 48 hours. Melts were then lowered to 50º C above the liquidus and water quenched. Glass transition temperatures, $T_g(x)$ of the ternary, and non-reversing relaxation enthalpy, $\Delta H_{nr}(x)$ , were established using a model 2920 MDSC from TA instruments. Measurements on fresh (3 weeks) and aged (3 and 5 months) samples relaxed at 300K were performed. We find $T_g(x)$ to increase monotonically with x in the $0 < x < 0.25$ range ( Fig.1a). Variations in $\Delta H_{nr}(x)$ show (Fig. 1b) a global minima in the $0.09 < x < 0.145$ range that gets sharper and deeper as glasses outside the window *age* at 300K. In the $0.20 < x < 0.23$ range, variations in $T_g(x)$ and $\Delta H_{nr}(x)$ show a mild glitch ( Fig.1a) and a satellite window (fig.1b) respectively. Raman scattering on glasses excited in the IR ( 1.02 μm ) were performed in a back scattering geometry using a Nicolet FT Raman module with model 670 FTIR bench at 1 cm$^{-1}$ resolution. Fourteen bands were identified ( Fig 2a) and their strengths were traced as functions of composition in order to monitor the nature of the molecular clusters and the degree of nanoscale phase separation, enabling identification of P- centered pyramidal (PYR), quasi-tetrahedral(QT) , ethylene-like (ETH) and $P_4Se_3$ cages, as well as Ge-centered corner-sharing (CS) and edge-sharing (ES). For example, concentration of QT Se=P(Se$_{1/2}$)$_3$ units display (Fig. 2b)





a global maximum near x = 0.09, in harmony with earlier $^{31}$P NMR measurements[15].
These considerations lead to the construction of a full ternary phase diagram showing the
regimes of the three generic elastic phases observed near the stiffness transition in these
alloys; details will be published elsewhere[16].

The central result of the present work is the observation of a deep and wide *reversibility
window* in the 0.09 < x < 0.145, or 2.27 < $r$ < 2.44 range; the depth of the window varies
dramatically with aging ( Fig. 2b). The window fixes the *three elastic phases*; *floppy* at $r$
< 2.27 , *intermediate* in the 2.27 < $r$ < 2.44 range (table 1), and *stressed rigid* at $r$ >
2.44 in the present ternary. Here $r$ = 2 +3x (ref.2). The *reversibility window deepens
relative to* compositions outside the *window* because window compositions do not age,
while those outside the window age. Floppy glasses age over a 3-month waiting period,
while *stressed-rigid* ones age over a 5-month period. The somewhat slower kinetics of
aging of the latter is incidentally due to their higher $T_g$s. Note the complete absence of
aging for glasses in the *reversibility window* even after a 5 - month waiting period.

The local structures  populated in the *reversibility window*  include CS ( $r$ = 2.40-2.67)
and ES ( $r$ = 2.67) Ge(Se$_{1/2}$)$_4$ tetrahedra, pyramidal ( $r$ = 2.40) P(Se$_{1/2}$)$_3$ and quasi-
tetrahedral ( $r$ = 2.28) Se=P(Se$_{1/2}$)$_3$ units. A count of Lagrangian constraints/atom (due to
bond-stretching and bond-bending forces) for each of these local structures[2] equals 3, the
degrees of freedom/atom, and highlights their *isostatically* rigid nature. The exceptional
thermal and elastic behavior of glasses in the *reversibility window* derives from the
isostatically rigid nature of their backbones both at local and intermediate ranges[7].  It is





plausible that the backbones of these alloys are composed of such isostatic units, as the width of the *reversibility window*[2,3,5,9,17] spans (table 1) a range of chemical stoichiometries that encompasses those of the isostatic local structures identified above. Thus, for example, the window begins near $r = 2.28$ where the concentration of QT unit ($r = 2.28$) is a maximum (Fig.2b), and the window ends near $r = 2.44$ where concentrations of PYR units ($r = 2.40$), CS ($r = 2.40$-2.67) and ES units ($r = 2.67$) is high. And as one would expect, molecular packing of these units as manifested in molar volumes of the glasses show (Fig.3) absence (presence) of aging effects for compositions in (outside) the reversibility window. Note that in the isostatic window the molar volume is nearly independent of coordination number.

In conclusion, intermediate phases are a general feature of network glasses. Here we have shown that a distinctive property of glasses in intermediate phases is that they do not age. The latter behavior is consistent with glasses in intermediate phases as being stress-free in character. A common classification of glass types is in terms of the temperature dependence of the viscosity of the supercooled melt, namely whether they exhibit a constant Arrhenius activation energy, or whether this energy increases as the melt is supercooled; the former materials are said to form *strong* glasses, the latter *fragile* ones[18]. It appears that glass compositions in the *windows* are *strong* and *do not age*, while those outside the windows are *fragile* that evolve in time, or age.

The parallels between reversibility and aging near the stiffness transition of the network backbone are most suggestive of an analogy with protein folding postulated in recent





skeletal models of living polypeptide chains.  There it was shown[13] that 26 diverse

proteins undergo a mechanical stiffness transition very similar to that in singly bonded

network glasses near the same average coordination number ($r$ = 2.41) found in the

glasses[13].  The width of the protein window $\Delta r$ = 0.03, may be compared to $\Delta r$ = 0.16

for the glasses studied here (table 1).  The width of the windows may be determined by

the strengths of residual interactions relative to the covalent constraints.  Here these

residual fluctuations may be the differences between P-pyramids and Ge- tetrahedra,

while in the proteins the important residual interactions involve H bonds[13].  Note that a

very narrow window centered at  $r$ = 2.33 of width $\Delta r$ =0.01 has been observed in Ge-S-I

glasses (table 1); this width may reflect the weakness of non-bonded I-I van der Waals

interactions[19]. Of course, protein functionality requires almost no aging and nearly

complete reversibility during the life of the protein; thus the reversibility window could

also be called the window of life.

*Thermal reversibility* and *lack of aging* represent some of the generic network functions

shared by glasses and proteins that are a consequence of s*elf-organization* of these

disordered systems.  Self-organized networks exhibit quite different properties and

behavior from networks generated by toy models.  For example, most glasses display a

high degree of self-organization, which is why they do not crystallize even when slowly

quenched.  Glasses relax according to stretched exponentials, whereas it has been found

that scale-free toy networks grow logarithmically[20].  Here we have shown that the

conditions for formation of reversible functionality in glasses and proteins are similarly





distinctive and can be characterized by the mechanical properties of their elastic backbones.

We thank M. Mabry and B.Zuk of ThermoNicolet Inc.for the Raman measurements. LPTL is Unite Mixte de Recherche CNRS No 7600. This work is supported by NSF grant DMR-01-01808.

Captions

Fig.1a. $T_g(r)$ trends in Ge-Se ($\Diamond$) , P-Se ($\square$) and P-Ge-Se ($\bullet$) glasses. The thick black line shows the $T_g(r)$ prediction based on SAT. Inset shows concentration of homopolar bonds projected by SAT to account for the observed $T_g(r)$ trend. (b) Trends in $\Delta H_{nr}(x)$ in the $Ge_xP_xSe_{1-2x}$ ternary showing the reversibility window in the $0.09 < x < 0.145$; the latter gets deeper and sharper upon aging of glass samples at 300 K.

Fig. 2a. Raman scattering of a ternary glass at x = 0.10 showing modes of quasi-tetrahedral (QT) units ( 500 cm$^{-1}$) , ethylenelike (ETH) $P_2Se_3$ units( 375 cm$^{-1}$) , pyramidal (PYR) $P(Se_{1/2})_3$ units ( 330 cm$^{-1}$) , $Se_n$ chain mode (CM) at 250 cm$^{-1}$ and 140 cm$^{-1}$, corner-sharing (CS) and edge-sharing ES $Ge(Se_{1/2})_4$ units near 200 cm$^{-1}$ and 217cm$^{-1}$ respectively. (b) shows a plot of Raman scattering strength of QT mode normalized to the CM at 250 cm$^{-1}$ in open circles, while filled circles give concentrations of the QT units inferred from $^{31}$P NMR, reference 15. New vibrational modes of P-rich units appear at higher x and will be discussed in ref.16.

Fig. 3. Molar volumes of present glasses measured 2 months ($\bullet$) and 6 months ($\circ$) after water quench. Note aging effects occur for glass compositions outside the reversibility window but not inside the window.





Fig. 1 a and b

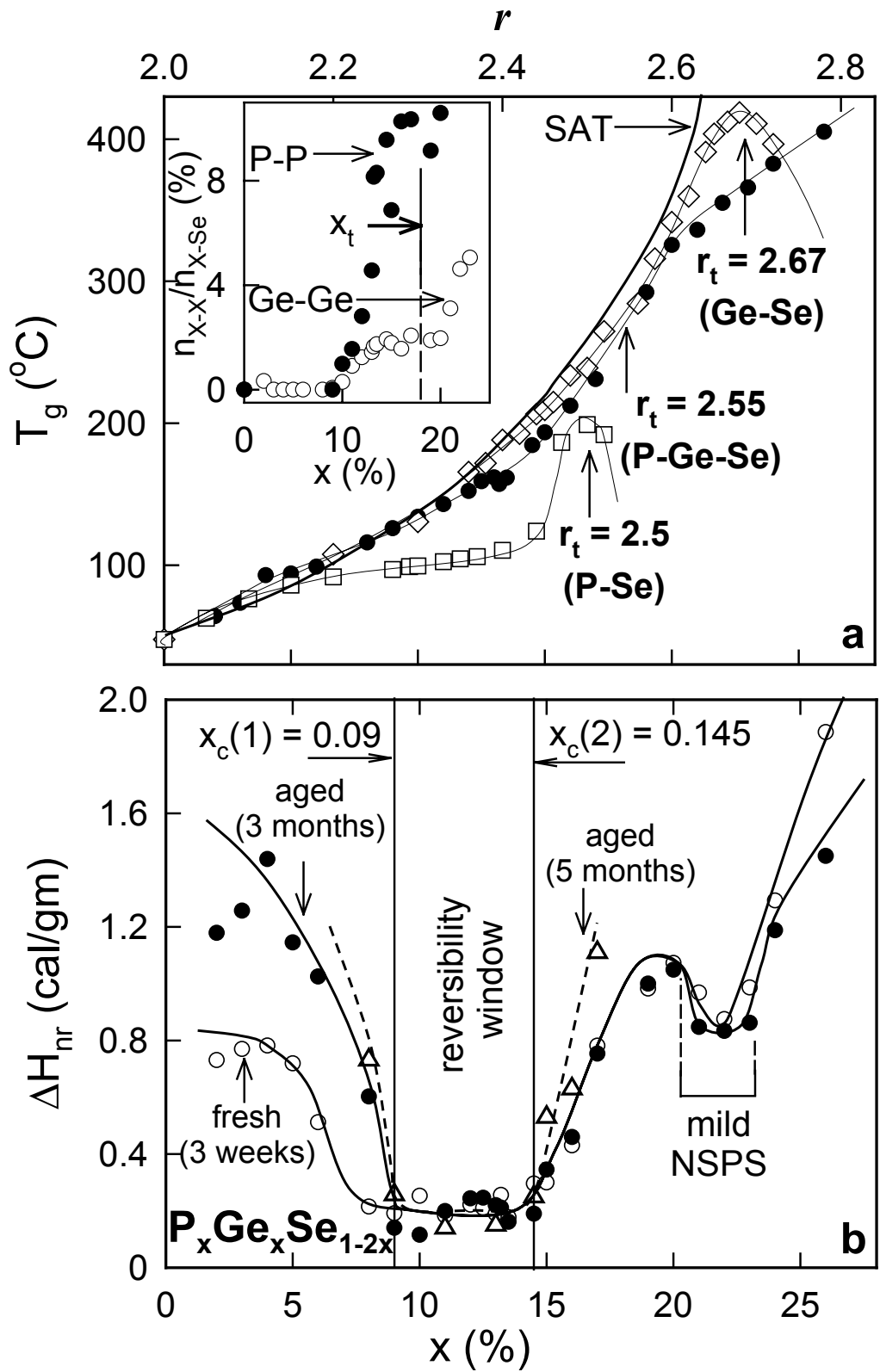





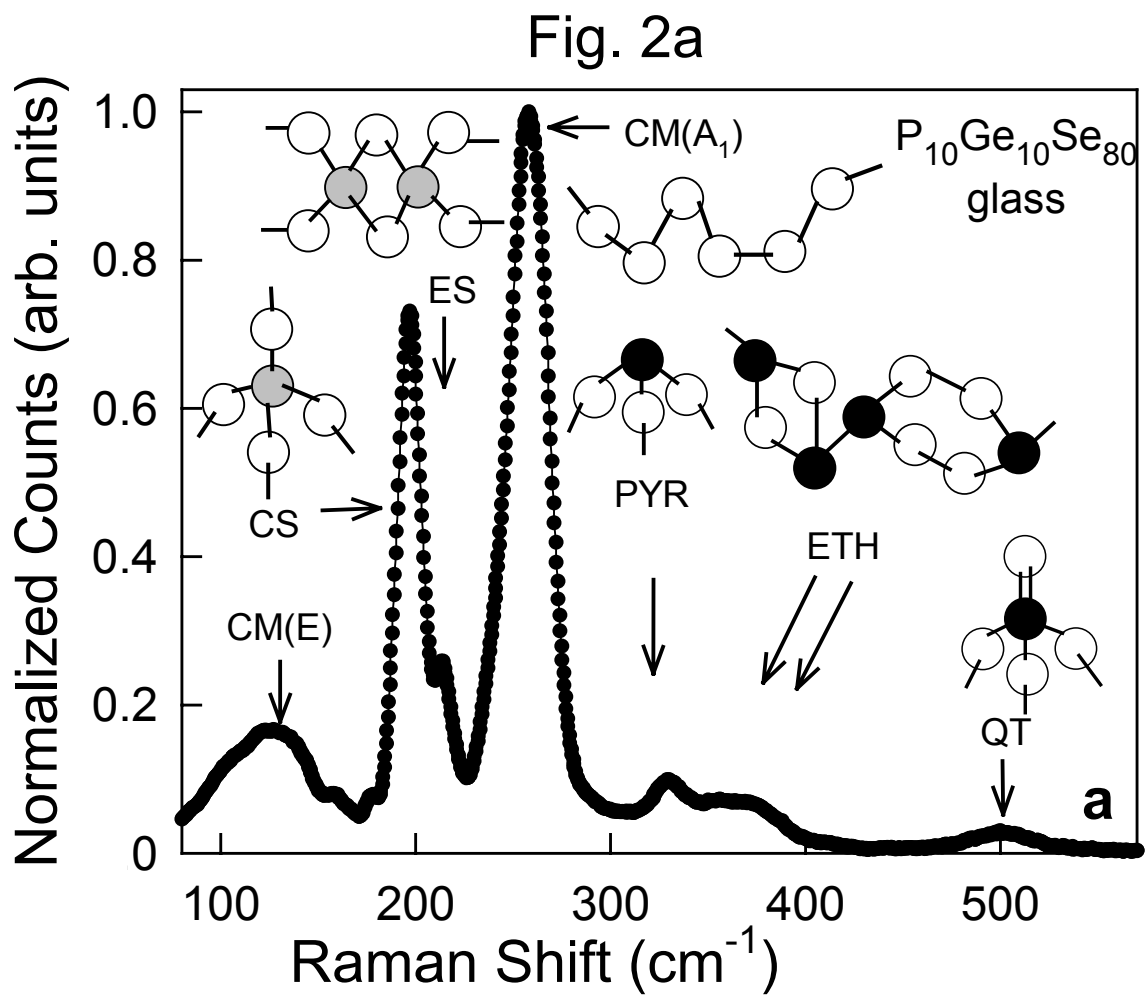

Fig. 2a





Fig. 2b

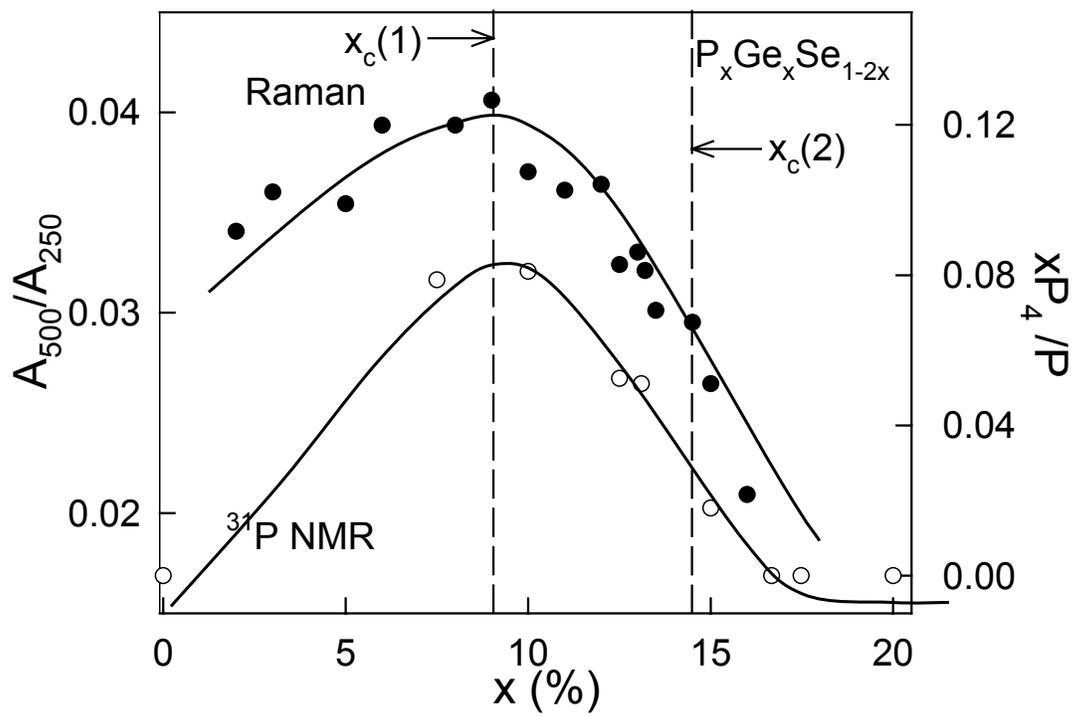





Fig. 3

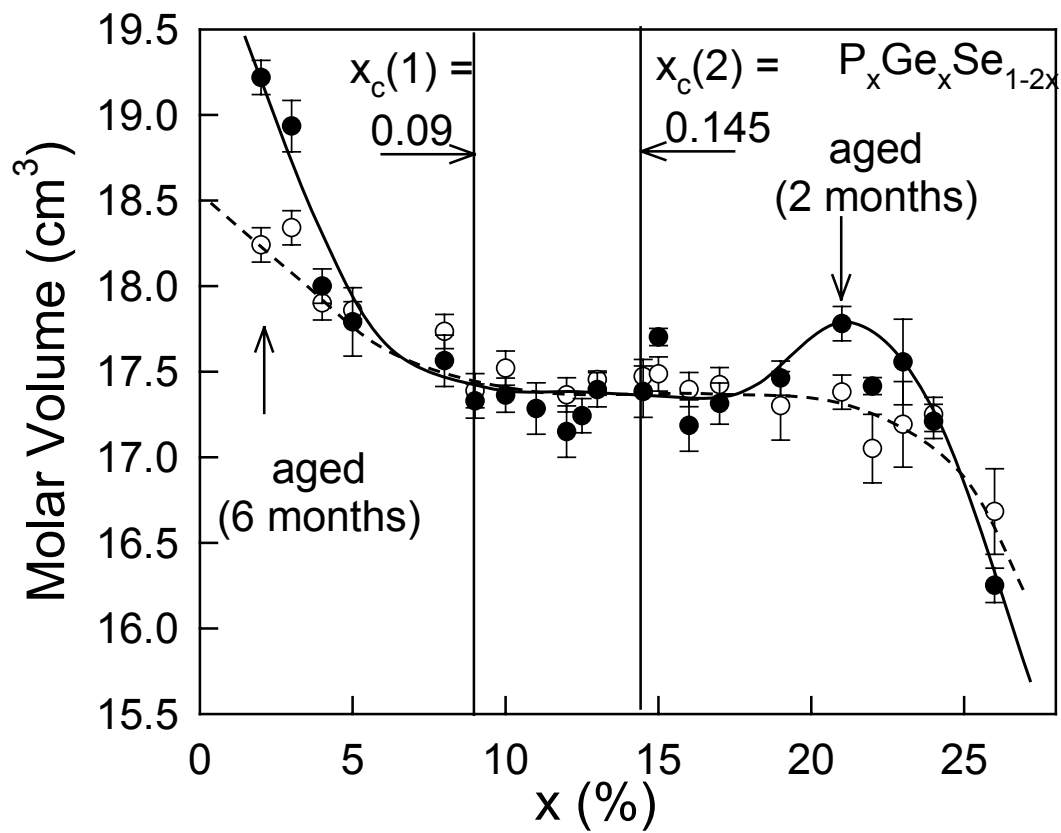





Table 1

| Network | Intermediate Phase | Ref. |
|---|---|---|
| | $r_1, r_2$ | |
| Ge-Se | 2.40, 2.52 | 3 |
| Si-Se | 2.40, 2.53 | 4 |
| As-Se | 2.29, 2.37 | 1 |
| P-Se | 2.28, 2.40 | 1 |
| Ge-S-I | 2.332, 2.342 | 19 |
| Ge-As-Se | 2.27, 2.46 | 5 |
| P-Ge-Se | 2.27, 2.43 | 16 |
| Proteins | 2.39, 2.42 | 13 |